\documentclass{PoS}
\usepackage{amsmath,amssymb,latexsym,amsbsy,bbm,makerobust}
\newcommand{\contains}{\ensuremath{\ni}}
\newcommand{\abs}[1]{\ensuremath{\left|#1\right|}}
\newcommand{\figref}[1]{Figure~\ref{fig:#1}}

\MakeRobustCommand\eqref

\title{Lattice QCD and Axion Cosmology}

\ShortTitle{Lattice QCD and Axion Cosmology}

\author{\speaker{Evan Berkowitz}\thanks{
Accompanying slides are available at \cite{slides}.
I am tremendously grateful to my collaborators on Ref.~\cite{BBR} Enrico Rinaldi and Michael Buchoff.
This work was performed under the auspices of the U.S. Department of Energy by LLNL under Contract No. DE-AC52-07NA27344. 
This research was partially supported by the LLNL Multiprogrammatic and Institutional Computing program through a Tier 1 Grand Challenge award.
\mbox{This is document LLNL-PROC-676578.}}\\
        Nuclear and Chemical Sciences Division, Physical and Life Sciences Directorate\\
        Lawrence Livermore National Laboratory\\
        E-mail: \email{berkowitz2@llnl.gov}}

\abstract{The Strong CP Problem can be resolved by introducing an additional global symmetry known as Peccei-Quinn symmetry.
Once PQ symmetry is broken the associated particle, the QCD axion, is a plausible dark matter candidate.  Calculating the cosmological energy density of the axion requires nonperturbative QCD input---the high-temperature topological susceptibility.  I will show results from a pure-glue calculation and examine the implications for the axion mass and coupling.}

\FullConference{The 33rd International Symposium on Lattice Field Theory\\
		14 -18 July 2015\\
		Kobe International Conference Center, Kobe, Japan*}

\begin{document}

\section{Introduction}

QCD could in principle violate CP, as it supports the so-called ``theta term'' in its Lagrangian:
\begin{equation}
	\mathcal{L}_{\text{QCD}} \contains \theta \frac{1}{32\pi^2} \epsilon^{\mu\nu\rho\sigma}F_{\mu\nu}F_{\rho\sigma}
\end{equation}
which violates CP thanks to symmetry properties of the Levi-Civita epsilon tensor.  This term is topological, in the sense that its spacetime integral is always an integer multiple of $\theta$,
\begin{equation}\label{Q}
	Q = 
	\frac{1}{32\pi^2}
	\int d^4x \;
	\epsilon^{\mu\nu\rho\sigma}F_{\mu\nu}F_{\rho\sigma}
	\in
	\mathbb{Z}
\end{equation}
and we call $Q$ the topological charge.
The parameter $\theta$ is truly an angle---it can in principle take any value only in $(-\pi,\pi]$, because the weight in the path integral
	$e^{i S_\text{QCD}} \propto e^{i Q \theta}$
is periodic in $\theta$, as $Q$ is an integer.
From limits on the neutron electric dipole moment
, an observable that should reflect any strong CP violation, one concludes that $\abs{\theta}\lesssim 10^{-10}$.
The Strong CP Problem asks: if $\theta$ could naturally take any value in $(-\pi,\pi]$, why is it so tiny?

Axions provide a dynamical mechanism to alleviate this fine tuning.
Peccei and Quinn observed that if the parameter $\theta$ were promoted to a dynamical field $a$ with the Lagrangian
\begin{equation}
	\mathcal{L}_{\text{axions}} = \frac{1}{2}\left(	\partial_\mu a	\right)^2 + \left(	\frac{a}{f_a} + \theta	\right) \frac{1}{32\pi^2} \epsilon^{\mu\nu\rho\sigma}F_{\mu\nu}F_{\rho\sigma}
\end{equation}
the Strong CP problem could be solved\cite{PQ}.
The vacuum expectation value of the axion field automatically adjusts itself so that $(a/f_a + \theta)$ vanishes, as can be seen from effective field theory arguments.
Moreover, the axion's mass $m_a$ satisfies the relation
\begin{equation}\label{axion-qcd}
	m_a^2f_a^2 
	=
	\left.\frac{\partial^2}{\partial \theta^2}F_{\text{QCD}}(T,\theta)\right|_{\theta=0} 
	=
	\chi(T)
\end{equation}
where $F_{\text{QCD}}(T,\theta)$ is the free energy of QCD, which depends on the temperature $T$ and the parameter $\theta$.  The relevant derivative of $F_{\text{QCD}}$ is the temperature-dependent topological susceptibility, denoted $\chi(T)$.
Note that $m_a$ is not a free parameter of axion physics: given $f_a$ and the QCD topological susceptibility we know $m_a$, which is temperature dependent.

\section{The Over-Closure Bound}
\label{sec:overclosure}

It was quickly realized that axions not only solve the Strong CP Problem but also are a plausible dark matter candidate\cite{Weinberg,Wilczek}.
So, it is interesting to use observational knowledge to try and constrain axion properties.
One such constraint, the over-closure bound, gives a lower bound on the axion mass by requiring that axions not account for more than the observed dark matter abundance\cite{AS}.

High temperature arguments imply that the QCD topological susceptibility $\chi$ vanishes at high temperatures, so that at the beginning of the universe axions are massless.
In particular, at asymptotically high temperatures the dilute instanton gas model (DIGM) should be valid, and it predicts
\begin{equation}\label{DIGM}
	\chi(T) = \frac{C}{(T/T_c)^n}
\end{equation}
where $T_c$ the temperature of the chiral transition can be traded, for example, for $\Lambda_\text{QCD}$ by simply readjusting $C$, and $n$ is predicted to be exactly $7$ when $N_c=3$.
We know from chiral perturbation theory ($\chi$PT), on the other hand, that at zero temperature the topological susceptibility is\cite{Weinberg}
\begin{equation}\label{chi0}
	\chi(0) = \frac{m_um_d}{(m_u+m_d)^2}f_\pi^2 m_\pi^2 \approx 3.6\cdot 10^7 \text{MeV}^4
\end{equation}
where $f_\pi \approx 93$MeV, so that $\chi$ must increase as temperature decreases, but only to a point.

As cosmological evolution proceeds the universe expands and cools.
From the equations of motion we can estimate that the axion number becomes fixed when roughly
\begin{equation}\label{T1}
	3H(T)\approx m_a(T)
	\hspace{1em}\text{or equivalently}\hspace{1em}
	9H^2(T) f_a^2 \approx \chi(T)
\end{equation} 
which fixes a temperature $T_1$ and a time $t_1$.
Put another way, as the universe evolves the topological susceptibility becomes large enough that the axion Compton wavelength is comparable to the size of the universe, and the axion's dynamics can kick in.
A more complete calculation would solve the equations of motion numerically, as is done in Reference~\cite{WS}.
We use
\begin{equation}
	H^2(T) = \frac{\pi^2}{90M_{\text{Pl}}^2}g_*(T)\; T^4
\end{equation}
where $M_\text{Pl}$ is the reduced Planck mass $\sqrt{\hbar c / 8 \pi G} \approx 2.4\times10^{18}$GeV and $g_*(T)$ is the effective number of degrees of freedom as a function of temperature given in Ref.~\cite{WS}.

After production ceases, axions follow a familiar but slightly modified story --- they get diluted by the universe's expansion but, unlike a typical dark matter candidate, continue to get more massive until the chiral transition of QCD freezes the topological susceptibility at today's value given in \eqref{chi0}.
An adiabatic analysis shows that\cite{Turner}
\begin{equation}\label{conserved}
	\frac{\rho(t)R(t)^3}{m_a(t)} = \text{the number of axions in a fixed comoving volume}
\end{equation}
is conserved instead of the usual $\rho(t)R(t)^3$.
The cosmological equations of motion provide the initial energy density\cite{PWW},
\begin{equation}\label{initial_condition}
	\rho(T_1) = \frac{1}{2} m_a^2f_a^2\theta_1^2 = \frac{1}{2} \chi(T_1) \theta_1^2
\end{equation}
which is a function of $\theta_1$---the quantity $\langle a \rangle /f_a + \theta$, the coefficient of the CP-violating term in the Lagrangian, at the temperature $T_1$.
If Peccei-Quinn breaking happens after inflation, then we can expect different regions of our universe to have different values of $\theta_1$.
Averaging over the different regions gives the value $\langle \theta_1^2 \rangle = \pi^2/3$.
From now on we will discuss only this late-breaking scenario---in a scenario where PQ symmetry breaks early, instead of a lower bound on $m_a$ we can get a constraint on a combination of $m_a$ and the initial condition $\theta_1$.

With mild algebraic massaging, spatially averaging $\theta_1$, and use of \eqref{axion-qcd}, we rewrite \eqref{conserved}
\begin{equation}\label{rho}
	\rho(T_\gamma) = \frac{\pi^2}{6}\sqrt{\chi(T_\gamma) \chi(T_1)}\left(\frac{R(T_1)}{R(T_\gamma)}\right)^3
\end{equation}
where $T_\gamma = 2.73$K is the present-day temperature of the CMB and where both factors clearly decrease with increasing $T_1$.
We use the $\chi$PT value $\chi(0)$ as an approximation for $\chi(T_\gamma)$.
Cosmology provides $R(T)$, so we now know the present-day axion abundance, given the topological susceptibility $\chi$ and the temperature $T_1$ (itself given by $\chi$, $f_a$, and cosmological inputs).
Since \eqref{T1} implies that for a fixed topological susceptibility $T_1$ decreases with increasing $f_a$, it is apparent that the axion energy density today increases with $f_a$, and that requiring $\rho(T_\gamma)$ be less than the observed dark matter energy density yields an upper bound on $f_a$ and subsequently a lower bound, through \eqref{chi0}, on the present-day axion mass.
Importantly, this argument also implies that if there is more than one kind of dark matter, the lower bound gets tighter.

Prior state-of-the-art calculations include calculations from the naive dilute instanton gas model (DIGM) and numerical calculations from the interacting instanton liquid model (IILM).
Reference~\cite{WS} found that the IILM predicts $m_a \geq 21 \pm 2\mu$eV.
However, at high temperatures they also found that various models differ in their predictions for $\chi(T)$ by an order of magnitude or more, so it is challenging to evaluate the systematic error in their result.
To make sure the axion is not mistakenly missed, the Axion Dark Matter Experiment (ADMX) takes a very conservative over-closure bound of $\sim1\mu$eV.
In principle, a calculation of $\chi$ directly from lattice QCD would yield totally controlled errors and an entirely trustable over-closure bound.

\section{Lattice Calculation}

\begin{figure}[ht]
	\centering
		\includegraphics[width=0.49\textwidth]{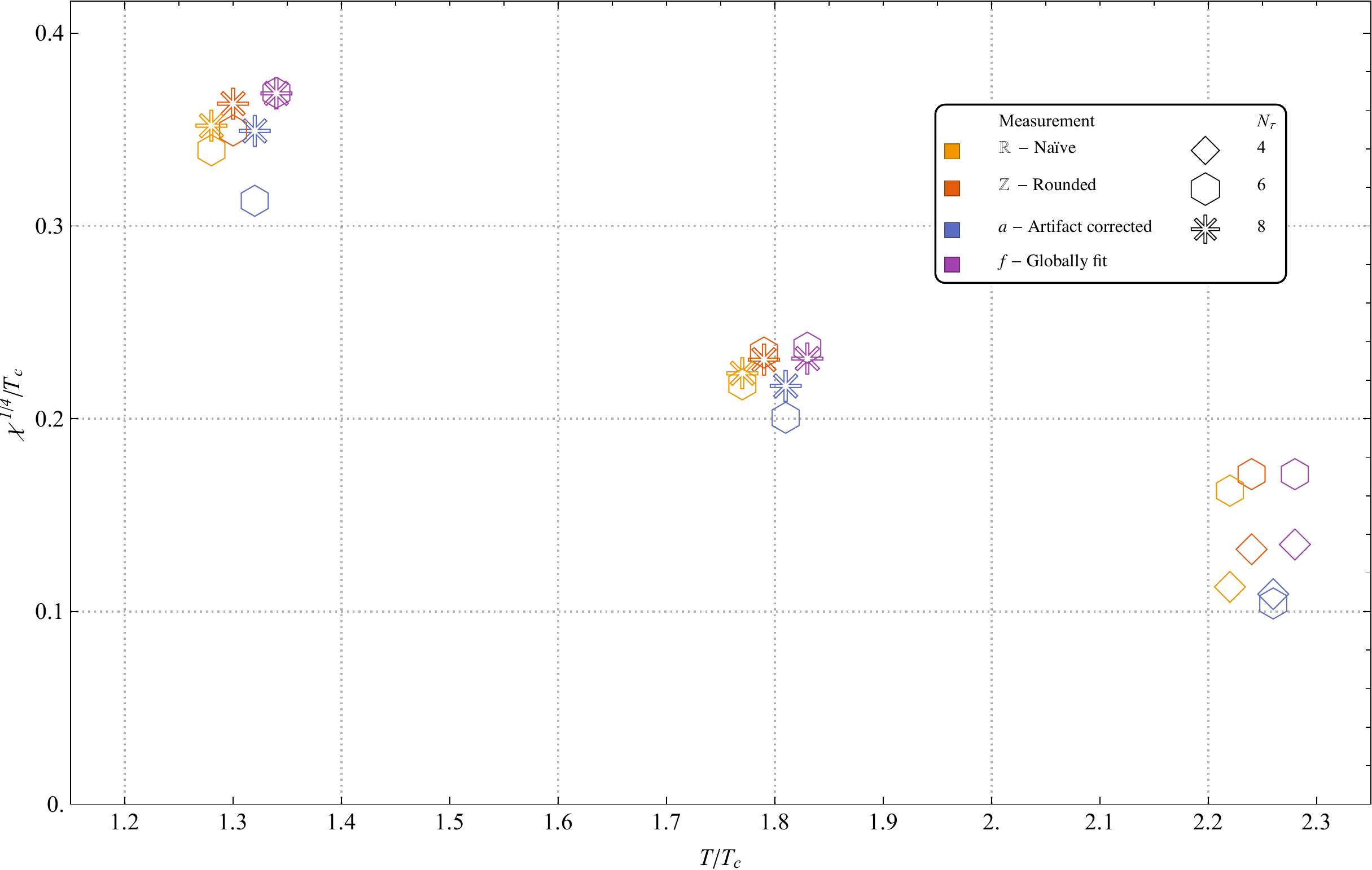}
		\includegraphics[width=0.49\textwidth]{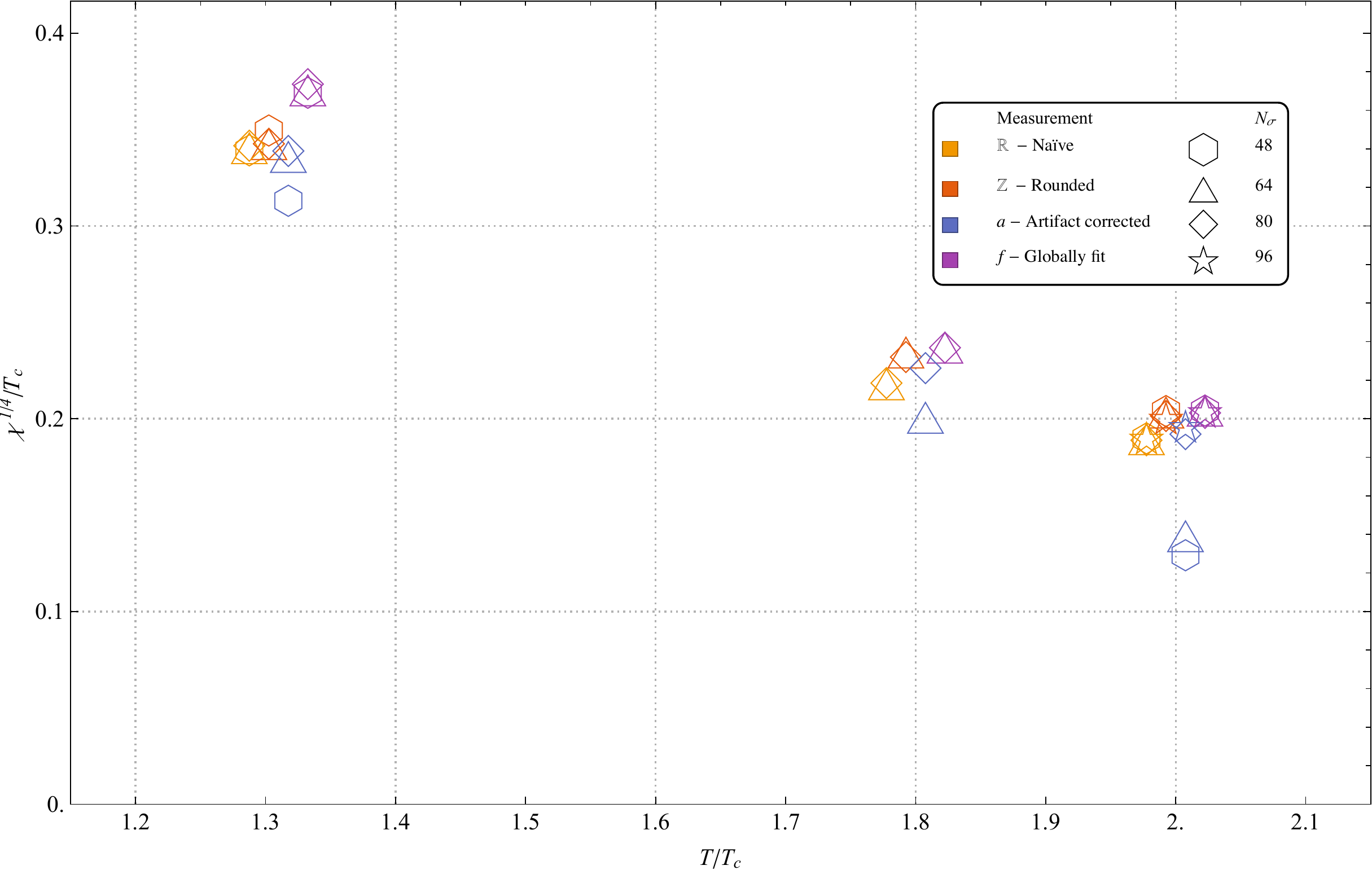}
	\caption{The left panel shows a study of discretization effects at three temperatures for all four definitions of the topological charge we use.  The right panel shows a study of finite volume effects at three different temperatures.  Each ensemble is slightly offset from its temperature for ease of visibility.  The statistical error bars on each data point are much smaller than the markers and are suppressed for clarity.}
	\label{fig:systematics}
\end{figure}

Unfortunately, a direct calculation from QCD will prove to be extremely expensive.
As a preliminary calculation, though, it makes sense to study pure glue---Yang--Mills with no quarks.
This simplification allows us to take advantage of heatbath algorithm, which enables us to calculate on ensembles with huge volumes and gather tremendous statistical samples with short autocorrelation times.
We studied temperatures in the range 1.2 to 2.5 times $T_c$ the critical temperature of pure glue  (approximately 284 MeV), at two different lattice spacings corresponding to $N_\tau$=6 and 8 time slices.
Previous studies went out to 1.3$T_c$, and our calculation agrees nicely with those results.
The ensembles are described in detail in Ref.~\cite{BBR}.
Other groups have since shown similar studies with different techniques\cite{KY,Mages}, and while there is some tension in the lattice results, even the most conservative (lowest) lower bound excludes all the parameter space that ADMX plans to search through 2017.
Additional discussion along these lines is postponed until the next section.

The topological susceptibility $\chi$ can be calculated on a finite volume by measuring the topological charge $Q$ on each configuration and taking the ensemble average,
\begin{equation}
	\chi = \lim_{V\rightarrow\infty}\frac{\langle Q^2 \rangle - \langle Q \rangle^2}{V}.
\end{equation}
We measure $Q$ through the naive discretized analogue of \eqref{Q},
\begin{equation}
	Q_\mathbb{R} = \frac{1}{32\pi^2} \sum_x \epsilon^{\mu\nu\rho\sigma}\square_{\mu\nu}\square_{\rho\sigma}
\end{equation}
where $\square$ is the lattice plaquette which reduces to the field strength $F$ in the continuum limit.
Because of the lattice discretization, $Q_\mathbb{R}$ is not an integer.
We massaged the distribution of $Q$ in four different ways (as detailed in Ref.~\cite{BBR}) and the method of Ref.~\cite{delDebbio} showed very little change as we changed the lattice spacing and size, as shown in \figref{systematics}.
Moreover, as the temperature increased the other definitions seemed to approach that definition, possibly indicating its reliability.
Thus, for everything that follows we used the results from that definition.

\begin{figure}[t]
	\centering
		\includegraphics[width=0.49\textwidth]{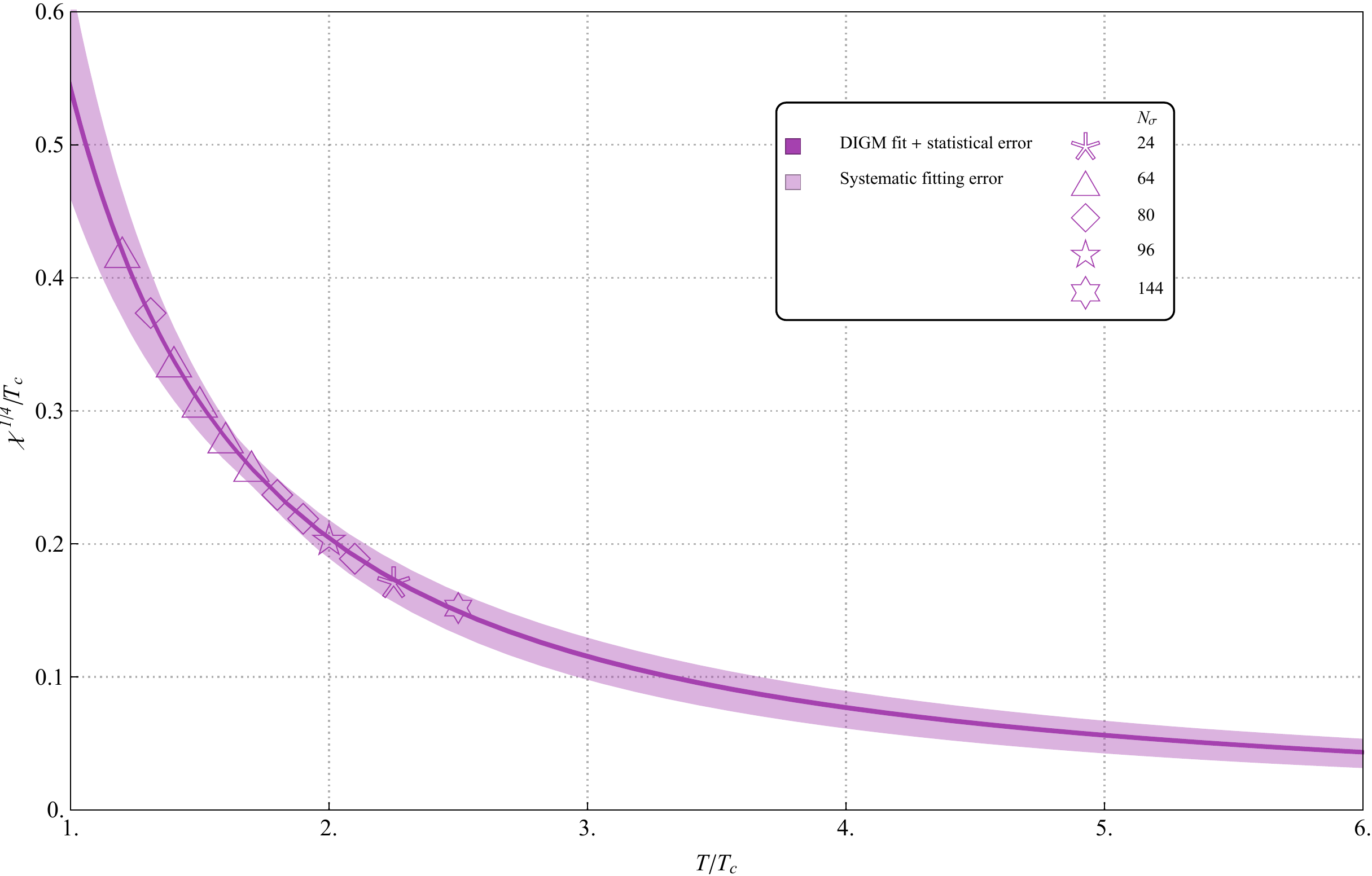}
		\includegraphics[width=0.49\textwidth]{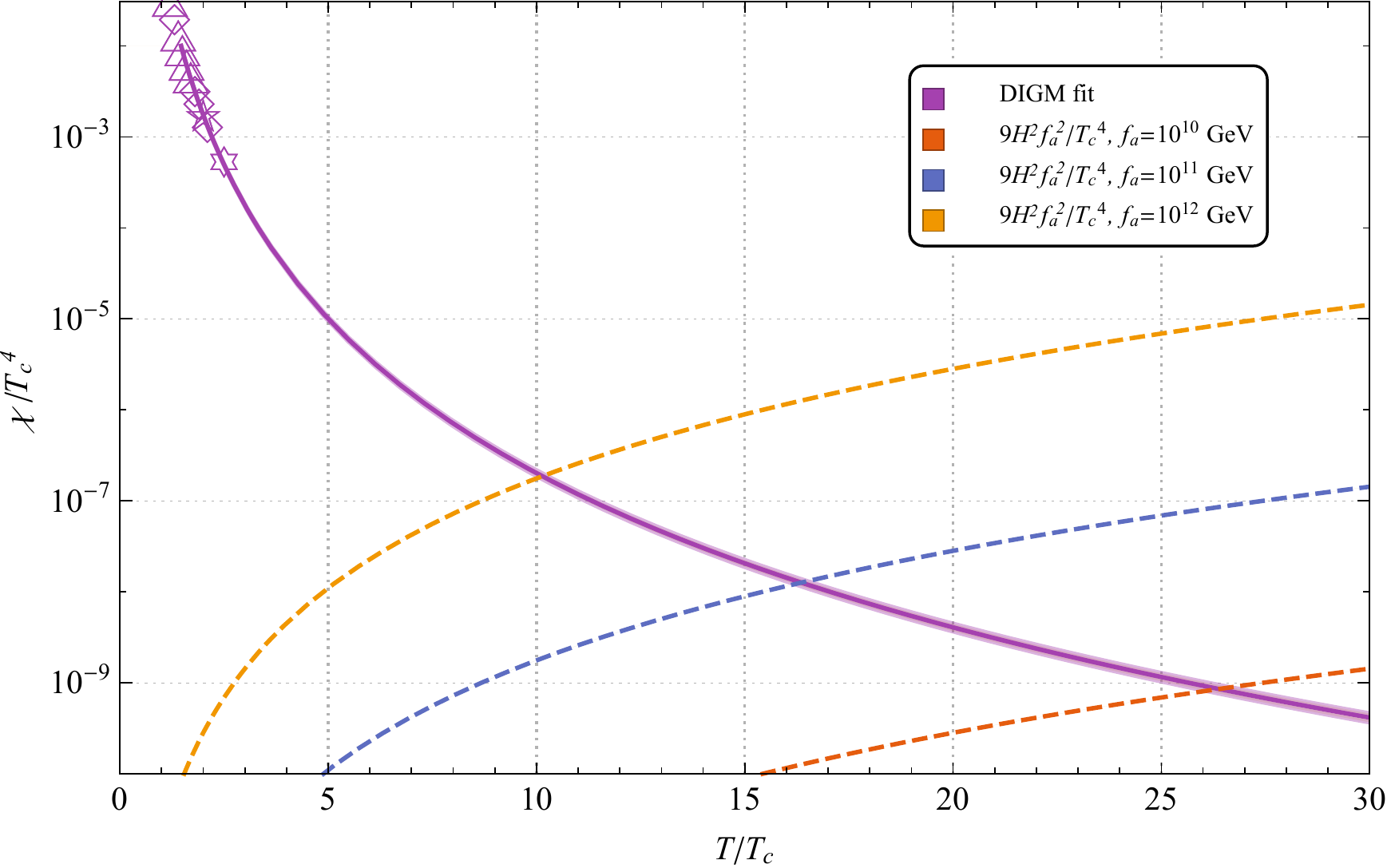}
	\caption{The left panel shows our best points with a DIGM-inspired (power-law) fit.
The right panel shows the two sides of \eqref{T1} for three example choices of $f_a$.  The intersection point for a given $f_a$ sets $T_1$.
The statistical error bars on each data point are much smaller than the markers and are suppressed for clarity.  
The statistical errors are thinner than the solid line---the band represents a very conservative systematic fitting error.  
More details can be found in Reference \cite{BBR}.}
	\label{fig:best}
\end{figure}

We did not study discretization effects at every temperature, so because we observe little lattice spacing dependence we simply take the largest volume at the coarser spacing, $N_\tau=6$.
Those points are shown in \figref{best}, with a fit inspired by the DIGM in \eqref{DIGM}, where we fit both the constant $C$ and the exponent $n$.
At asymptotically high temperatures $n=7$, but we find that, at these temperatures near $T_c$, $n=5.64\pm0.04$.
It is interesting to understand how this difference disappears and $\chi(T)$ is eventually described by a dilute instanton gas\cite{Mages}.

\section{Pure-Glue Bound}

We can use the best DIGM-inspired fit to extrapolate to high temperatures.
Of course, eventually it is important to have the region of cosmological interest under numerical control, especially if one desires a rigorous bound from QCD.

Plugging $\chi(T)$ given by our DIGM-inspired fit into \eqref{T1} and \eqref{rho} and requiring that $\rho(T_\gamma)$ accounts for at most the 27\% of the universe that is dark matter implies $f_a < (4.10\pm0.04)10^{11}$GeV and, through \eqref{chi0}, the present-day axion mass $m_a>(14.6\pm0.1)\mu$eV.
This is much higher than the bound ADMX uses for guidance, and excludes the masses ADMX plans to search through 2017.

While our calculation is not a calculation from full QCD, it might still be useful for experimental guidance, as will be explained now.
General arguments suggest that $\chi(T)$ should vanish if at least one quark mass vanishes.
Pure glue corresponds to the opposite limit, with all quark masses going to infinity.
It's extremely plausible, then, that the topological susceptibility for real QCD would be below that of pure glue.\footnote{Some speakers\cite{Cossu,Sharma} at LATTICE 2015 show values of $\chi(T)$ at $1 \lesssim T/Tc \lesssim 1.2$ for domain wall or overlap fermions with light pion masses.}

We further observe that with either a smaller or faller-fasting topological susceptibility the mass bound always gets stronger.
We can show this by considering a fixed $f_a$ which saturates the bound for a given $\chi(T)$ and asking how $\rho(T_\gamma)$ changes with a small change in the topological susceptiblity $\chi(T)\rightarrow\chi(T)+\delta\chi(T)$.  
If $\rho(T_\gamma)$ increases, the previously-allowed $f_a$ will be disallowed and the mass bound will be stronger.
Equation \eqref{T1} indicates $T_1$ shifts by $\delta T_1$ given by
\begin{equation}
	\frac{\delta T_1}{T_1} = \frac{1}{4 - \frac{T_1}{\chi(T_1)} \frac{\partial \chi}{\partial T}} \frac{\delta\chi(T_1)}{\chi(T_1)}
\end{equation}
(assuming $g_*(T_1)$ to be a constant) and thus from \eqref{rho} $\rho(T_\gamma)$ changes by an amount $\delta\rho$ satisfying
\begin{equation}\label{change}
	\frac{\delta\rho}{\rho} 
	= \frac{
		2 + 3 \frac{T_1}{R(T_1)} \left.\frac{\partial R}{\partial T}\right|_{T_1}
		}
		{
		4 - \frac{T_1}{\chi(T_1)} \left.\frac{\partial\chi}{\partial T} \right|_{T_1}
		}
		\;\;
		\frac{\delta\chi(T_1)}{\chi(T_1)}.
\end{equation}
Since the denominator in \eqref{change} is positive the sign of $\delta\rho$ is controlled by the numerator.
In the cosmological region of interest the universe is radiation-dominated and thus $R \propto T^{-1}$, so that $\delta\rho$ is positive when $\delta\chi$ is negative---decreasing $\chi(T)$ tightens the over-closure bound.

Since it's quite plausible that the topological susceptibility of real QCD is below that of pure glue, we expect the mass bound from QCD to be a stronger bound than our pure-glue bound of $m_a > (14.6\pm0.1)\mu$eV.
This strengthens the value of our bound, in that it suggests our bound is more conservative than the bound from QCD.

\section{Conclusions and Outlook}

Peccei--Quinn symmetery nicely cleans up the Strong CP problem and simultaneously provides a plausible, largely unconstrained dark matter candidate---the axion.
If the topological susceptibility of full QCD is indeed lower than that of pure glue, if the estimate \eqref{T1} is reliable, and if PQ symmetry broke after inflation, then with its current operational plan ADMX cannot find axions until 2018.
If other dark matter species account for a sizable portion of the total dark matter abundance, the over-closure bound will be tighter and the possibility of experimental observation of axions will be further postponed.
On the other hand, if an axion is discovered with a mass lower than our lower bound $\sim15\mu$eV, a complete calculation of the over-closure bound from QCD becomes urgent, because if that axion should be excluded (as our arguments in the previous section suggest) it implies that PQ symmetry broke early---a possible signal of preinflationary dynamics.

One future direction is the calculation of higher moments of the topological charge at finite temperature, which may prove necessary for full numerical control over the cosmological axion equations of motion.
Additionally, the incorporation of quarks and the measurement of the topological charge $Q$ via the Wilson flow method are obvious steps towards a fully controlled bound.
At the temperatures of cosmological interest, topological freezing may make it necessary to measure the susceptiblity $\chi$ using other techniques.
Obvious candidates include fixed topology methods and open boundary conditions.
Finally, QCD with real, finite $\theta$ may be amenable to Langevin methods, which would permit a full calculation of the free energy $F_\text{QCD}(T,\theta)$.

\end{document}